\documentclass[reprint,prb,twocolumn,showpacs,superscriptaddress,aps]{revtex4-1}

\usepackage{graphicx}
\DeclareGraphicsExtensions{.png,.jpg,.eps}

\usepackage{float}
\usepackage{xcolor}
\usepackage{amsmath}
\usepackage{caption}
\usepackage{subcaption}
\usepackage{float}

\usepackage{wrapfig}

\begin{document}

\title{Ferromagnetic and insulating behavior of LaCoO$_3$ films grown on a (001) SrTiO$_3$ substrate. A simple ionic picture explained ab initio.}

\author{Adolfo O. Fumega}
  \email{adolfo.otero@rai.usc.es}
\author{V. Pardo}
  \email{victor.pardo@usc.es}
\affiliation{Departamento de F\'{i}sica Aplicada,
  Universidade de Santiago de Compostela, E-15782 Campus Sur s/n,
  Santiago de Compostela, Spain}
\affiliation{Instituto de Investigaci\'{o}ns Tecnol\'{o}xicas,
  Universidade de Santiago de Compostela, E-15782 Campus Sur s/n,
  Santiago de Compostela, Spain}

%\date{\today} 

\begin{abstract} 
This paper shows that the oxygen vacancies observed experimentally in thin films of LaCoO$_3$ subject to tensile strain are thermodynamically stable according to ab initio calculations. By using DFT calculations, we show that oxygen vacancies on the order of 6 \% 
forming chains perpendicular to the (001) direction are more stable than the stoichiometric solution. These lead to magnetic Co$^{2+}$ ions surrounding the vacancies that couple ferromagnetically. The remaining Co$^{3+}$ cations in an octahedral environment 
are non magnetic. The gap leading to a ferromagnetic insulating phase occurs naturally and we provide a simple ionic picture to explain the resulting electronic structure.

\end{abstract}

\maketitle

\section{Introduction}

Ferromagnetic insulators (FMI) and in particular ferromagnetic oxides are uncommon materials. The usual situation is that a ferromagnetic coupling occurs together with metallic behavior. 
In the case of oxides, the most common situation for insulators is that they order antiferromagnetically. 
However, several examples of ferromagnetic oxides do exist, such as spinels, \cite{Foncuberta_spinel,Zhang_spinel} perovskite-based La$_{1-x}$Sr$_x$MnO$_3$, \cite{Cesaria_LSMO} La$_2$CoMnO$_6$, \cite{Min_La2CoMnO6} etc. They are of extreme importance in the fabrication of spintronic devices. \cite{Fert_spintronics,gotte_spintronics,PRB_Michetti,ferromagnetic_topological_insulator_Katmis}
Due to their technological relevance and rarity, it is interesting to explore and clarify the properties of one of such systems since this could help in designing other oxides with improved properties.

LaCoO$_3$ (LCO) is a very intriguing insulator with a not-completely understood spin-state transition, \cite{SENARISRODRIGUEZ1995224,PhysRevLett.97.176405,doi:10.1143/JPSJ.67.290,PhysRevB.47.16124,PhysRevB.67.224423} and also an interesting excitonic behavior. \cite{PRB_Juan_Excitonic_LCO}  
In its bulk form  it presents a diamagnetic response at low temperatures, all the Co$^{3+}$ cations being in a non-magnetic low-spin state, with its all 6 electrons occupying the lower-lying $t_{2g}$ levels. 
As temperature is increased, a crossover to other spin states occurs leading to an increase in the magnetic susceptibility. However, when LCO is grown as a thin film on a substrate that induces epitaxial strain on it, a ferromagnetic phase 
arises at low temperatures, characterized by a saturation magnetization close to $0.8$ $\mu_B/Co$, in case the substrate is SrTiO$_3$ (STO), and a T$_c$ of about $80$ $K$. \cite{PRB_MVarela,Fran_PAD_LCO_STO} Early explanations \cite{LCO_spinstate_Blaha_Wentzcovitch} discussed this magnetic ordering as arising in a stoichiometric strained form of LCO from a spin-state ordering formed by a Co$^{3+}$ HS-LS alternating pattern. However, more recent experiments have shown that epitaxial strain induces the formation of oxygen vacancy planes that could induce the appearance of magnetic moments that could eventually couple ferromagnetically. \cite{PRB_MVarela,PRL_MVarela,doi:10.1021/nl1034896} 

In Ref. \onlinecite{PRL_MVarela} a combination of atomically resolved Z-contrast imaging and electron-energy-loss spectroscopy is used to prove that, when LCO is grown on STO (001), oxygen vancancy 
planes are formed perpendicular to the substrate with a repetition of one vacancy plane every three unit cells. This pattern has also been interpreted as coming from a kind of spin-state ordering providing the contrast.\cite{doi:10.1021/nl302562f}

Taking those images as a basis for our calculations, here we present a DFT study on the possible oxygen-vacancy configurations consistent with the Z-contrast imaging pictures published by various groups in different samples, grown by physical methods such as pulsed laser deposition\cite{PRB_MVarela,PRL_MVarela} or by chemical methods such as polymer assisted deposition\cite{Fran_PAD_LCO_STO} that could account for the ferromagnetic insulating behavior found experimentally. Our total energy calculations in the various possible configurations allow to understand the amount of vacancies that are stable, what environments for the neighboring cations these produce, how the local moments couple and the resulting electronic structure. 
Thus, our work helps clarifying the origin of the insulating behavior found in these thin films.

\begin{figure}[!hb]
\begin{center}
\includegraphics[width=0.7\columnwidth]{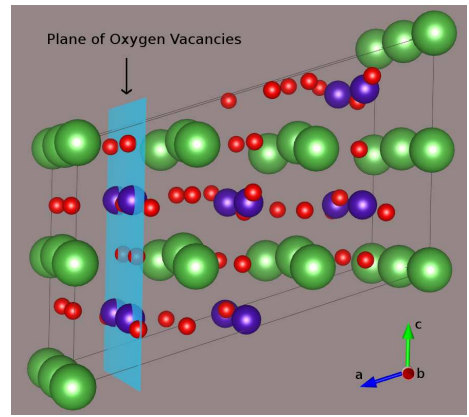}
\end{center}
\caption{(Color online.) Unit cell representation of La$_{12}$Co$_{12}$O$_{36-n}$ used in the calculations. La in green, Co in purple and O in red. The blue plane corresponds to the plane 
in which the oxygen vacacies are imposed (Fig. \ref{O_vac_plane_pos} shows the different configurations considered for the oxygen vacancies). }
\label{unit_cell}
\end{figure}

\begin{figure}[H]
\begin{center}
\includegraphics[width=0.42\columnwidth]{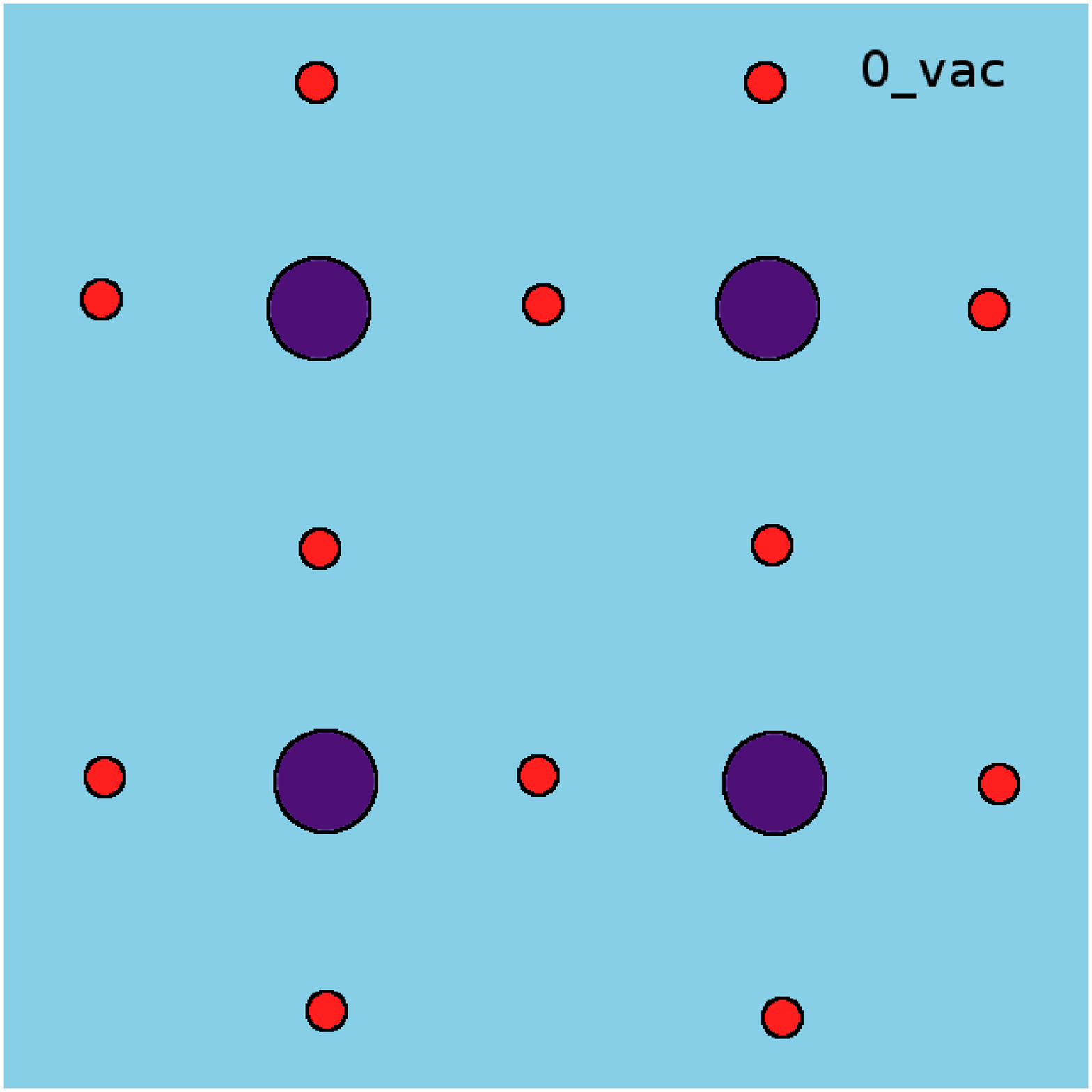}
\includegraphics[width=0.42\columnwidth]{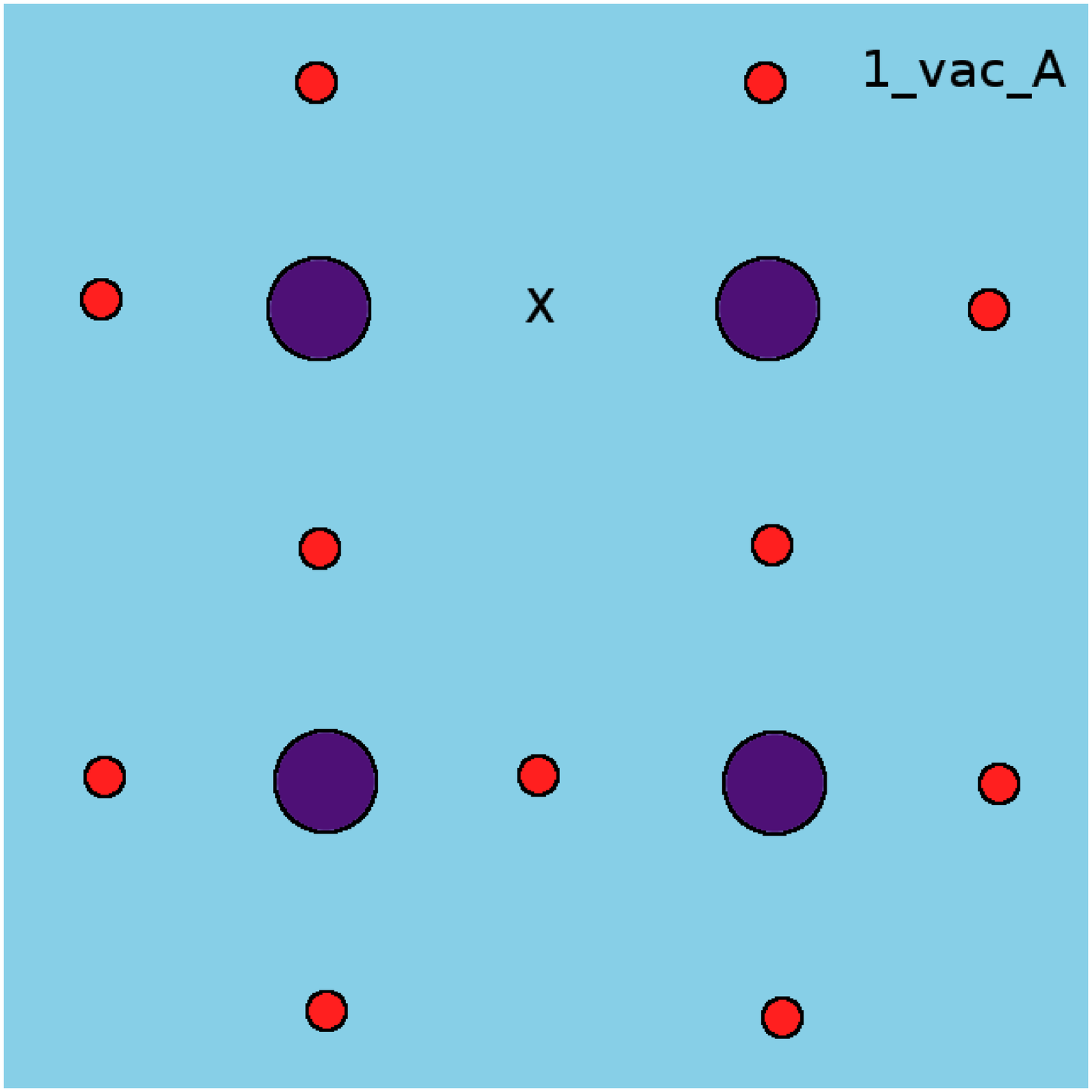}
\includegraphics[width=0.42\columnwidth]{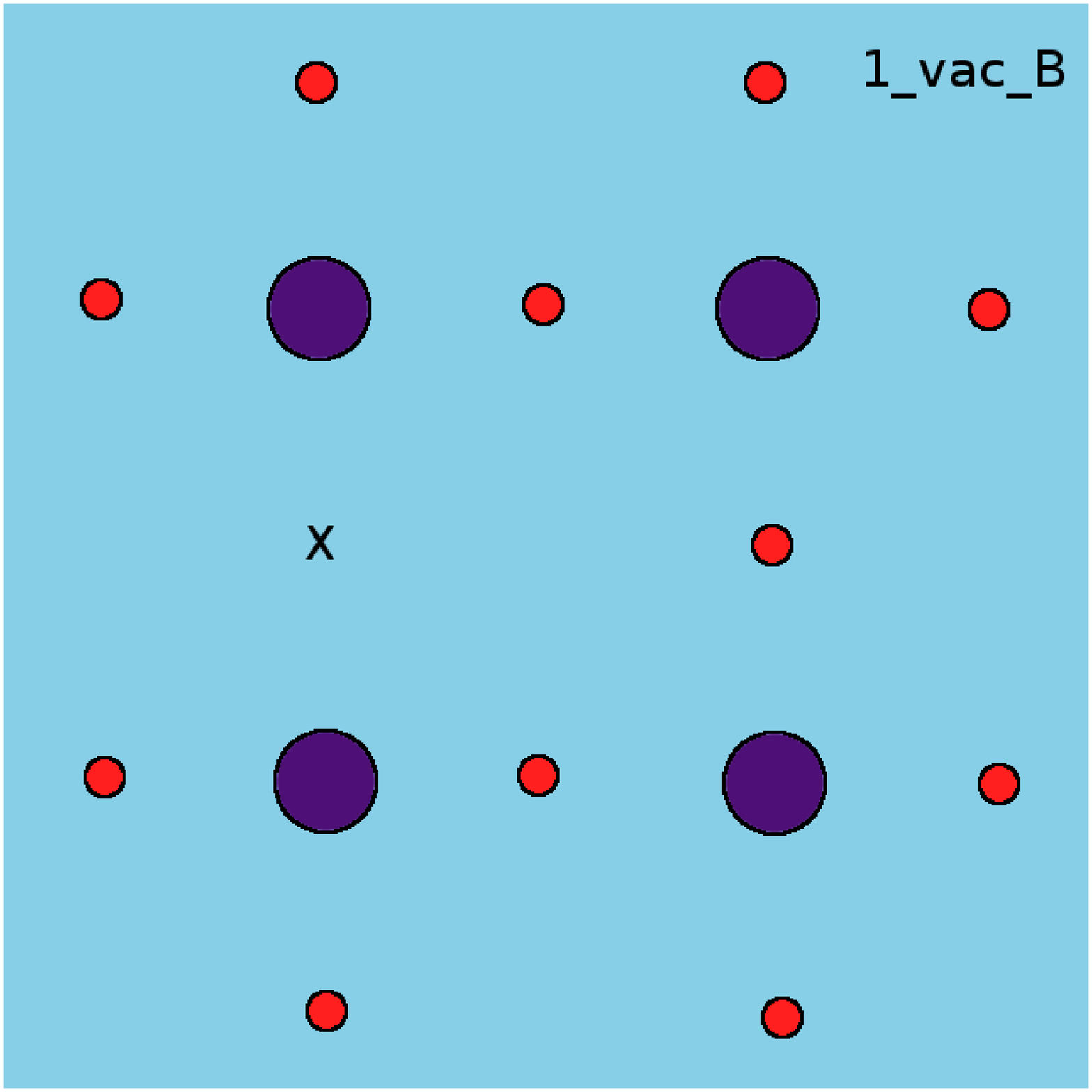}
\includegraphics[width=0.42\columnwidth]{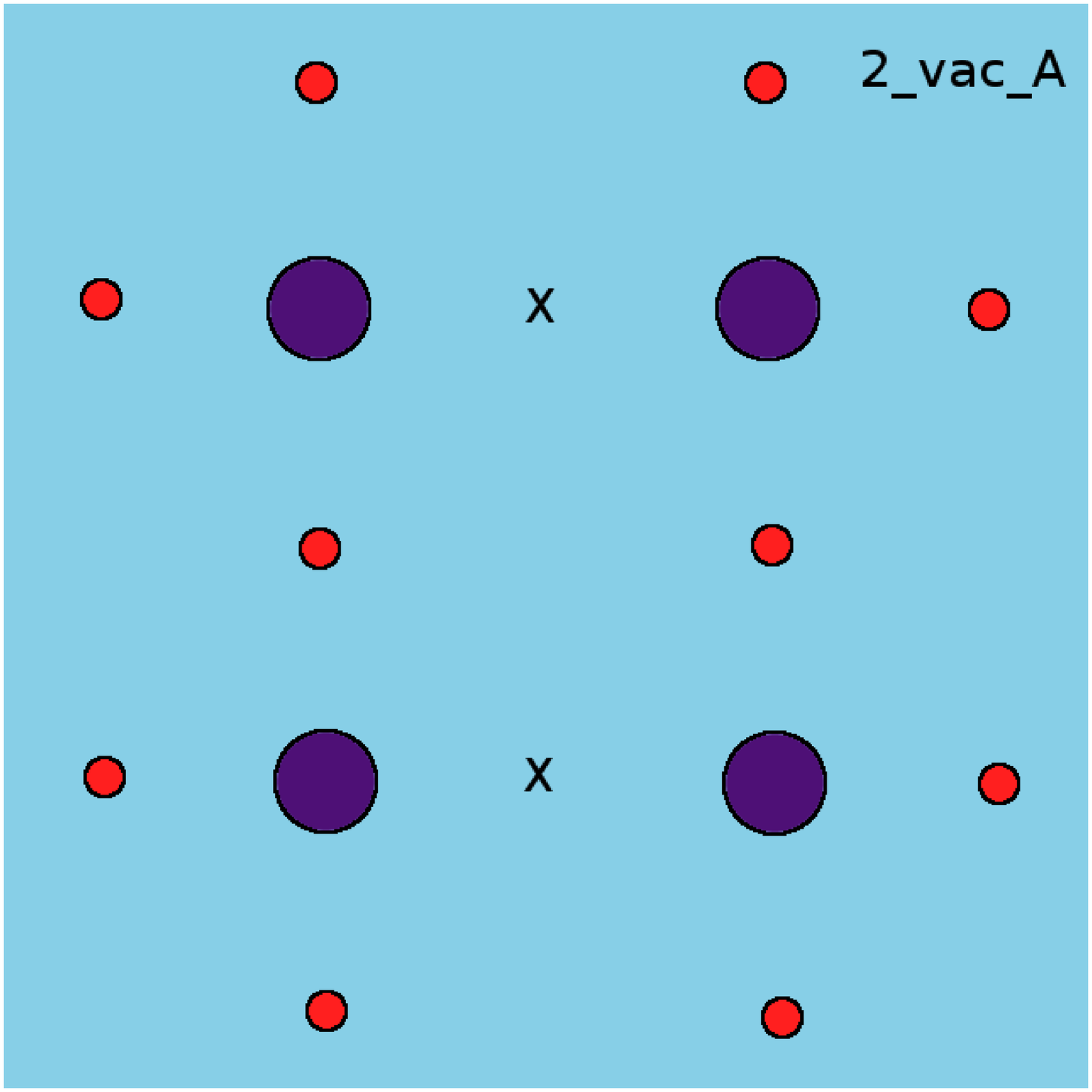}
\includegraphics[width=0.42\columnwidth]{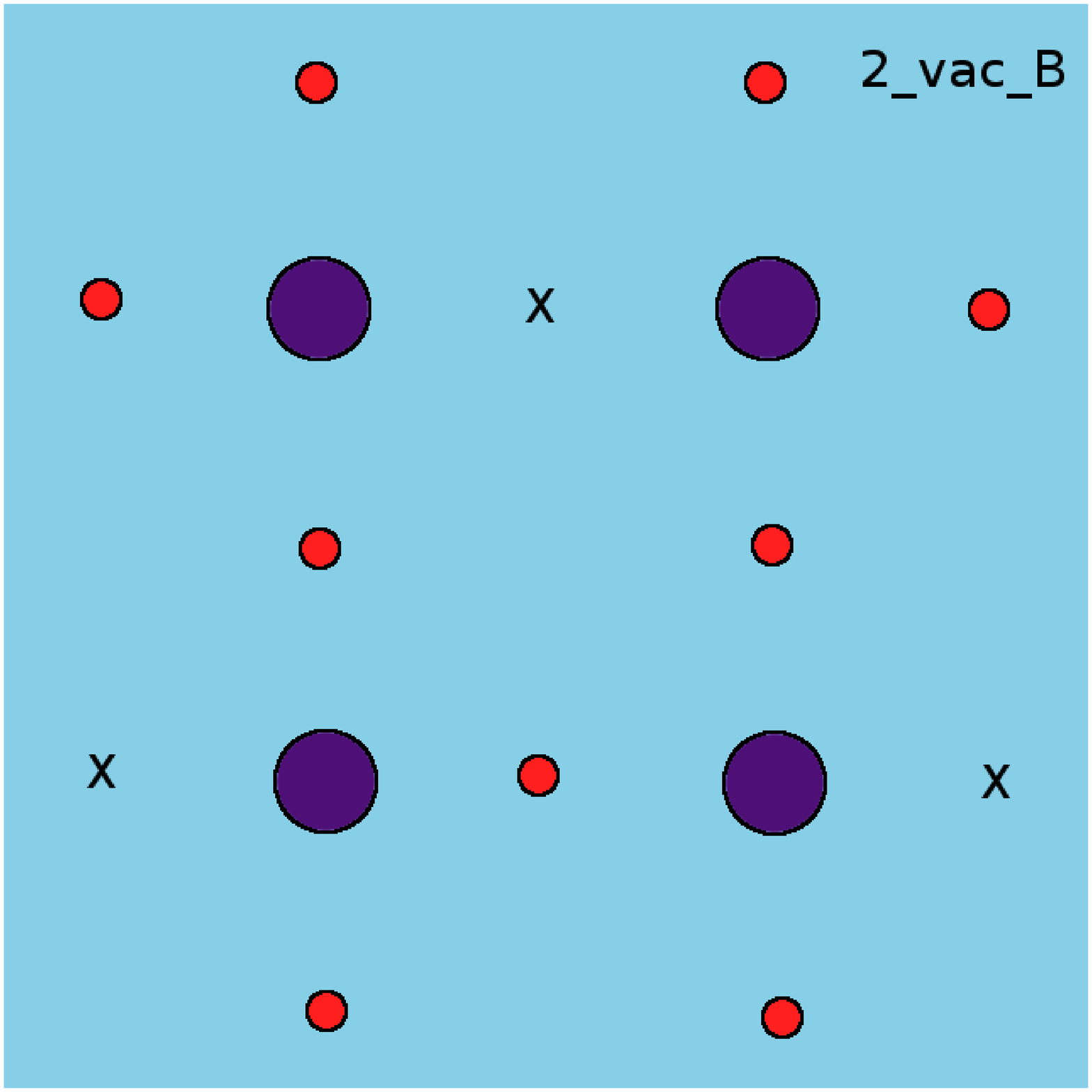}
\includegraphics[width=0.42\columnwidth]{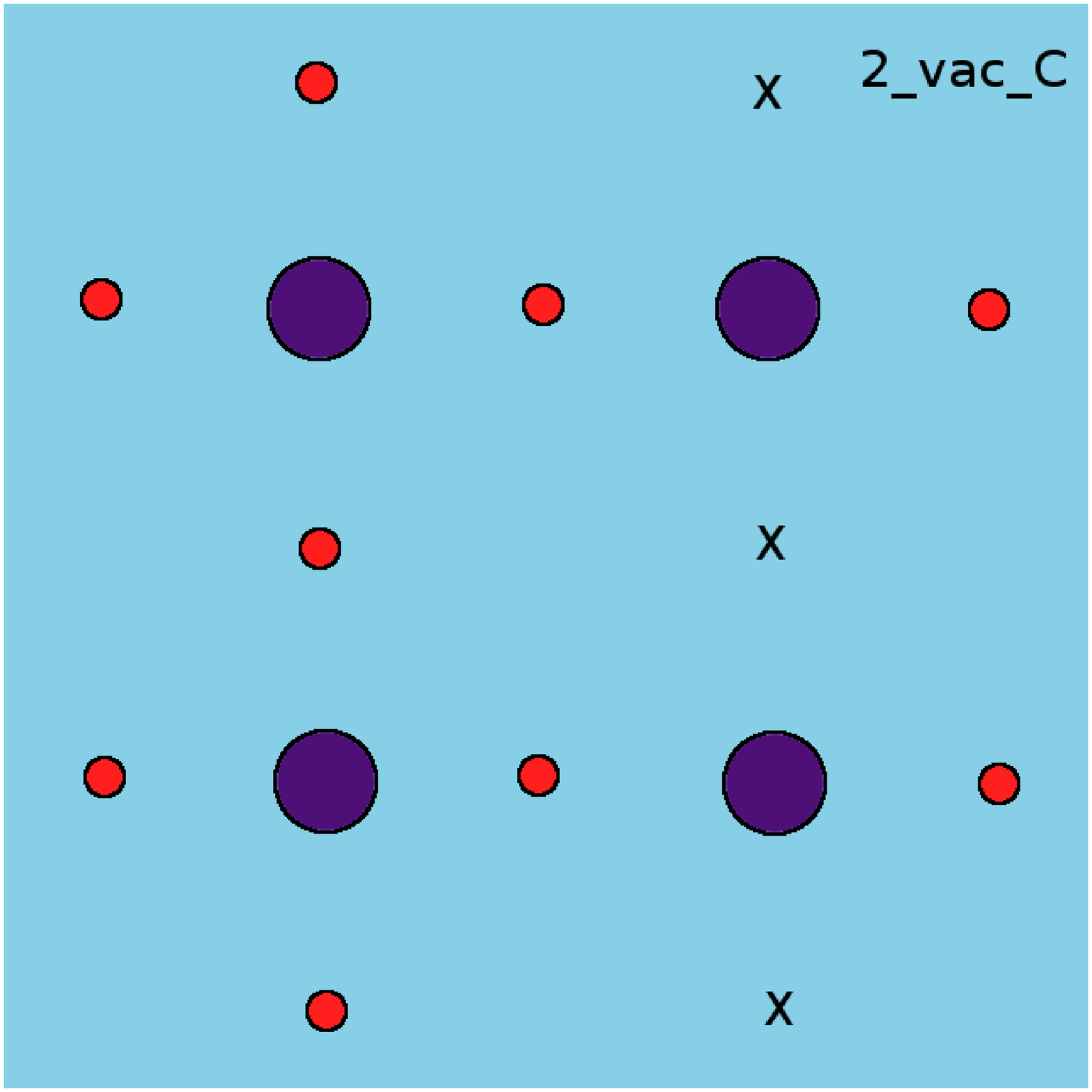}
\includegraphics[width=0.42\columnwidth]{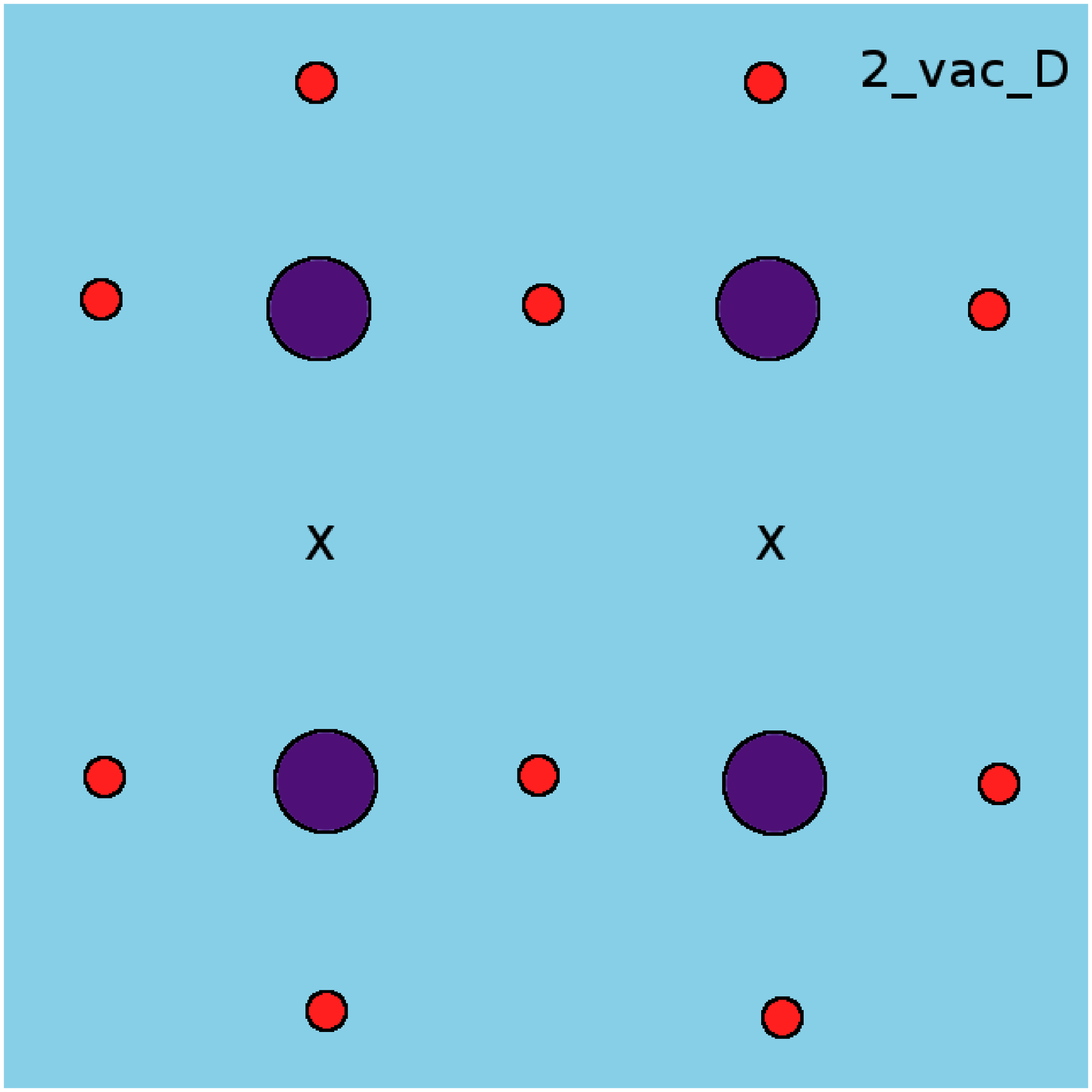}
\includegraphics[width=0.42\columnwidth]{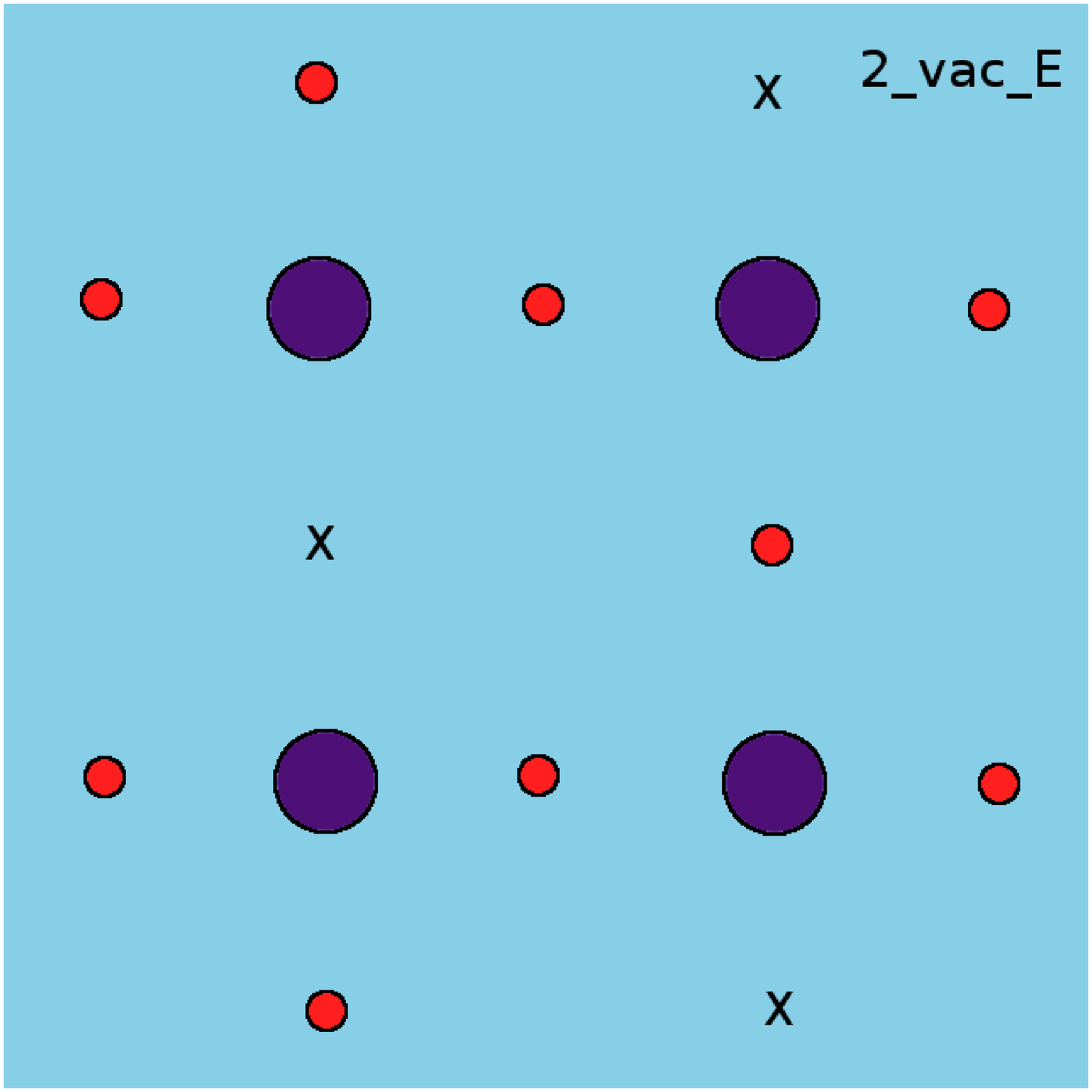}
\includegraphics[width=0.42\columnwidth]{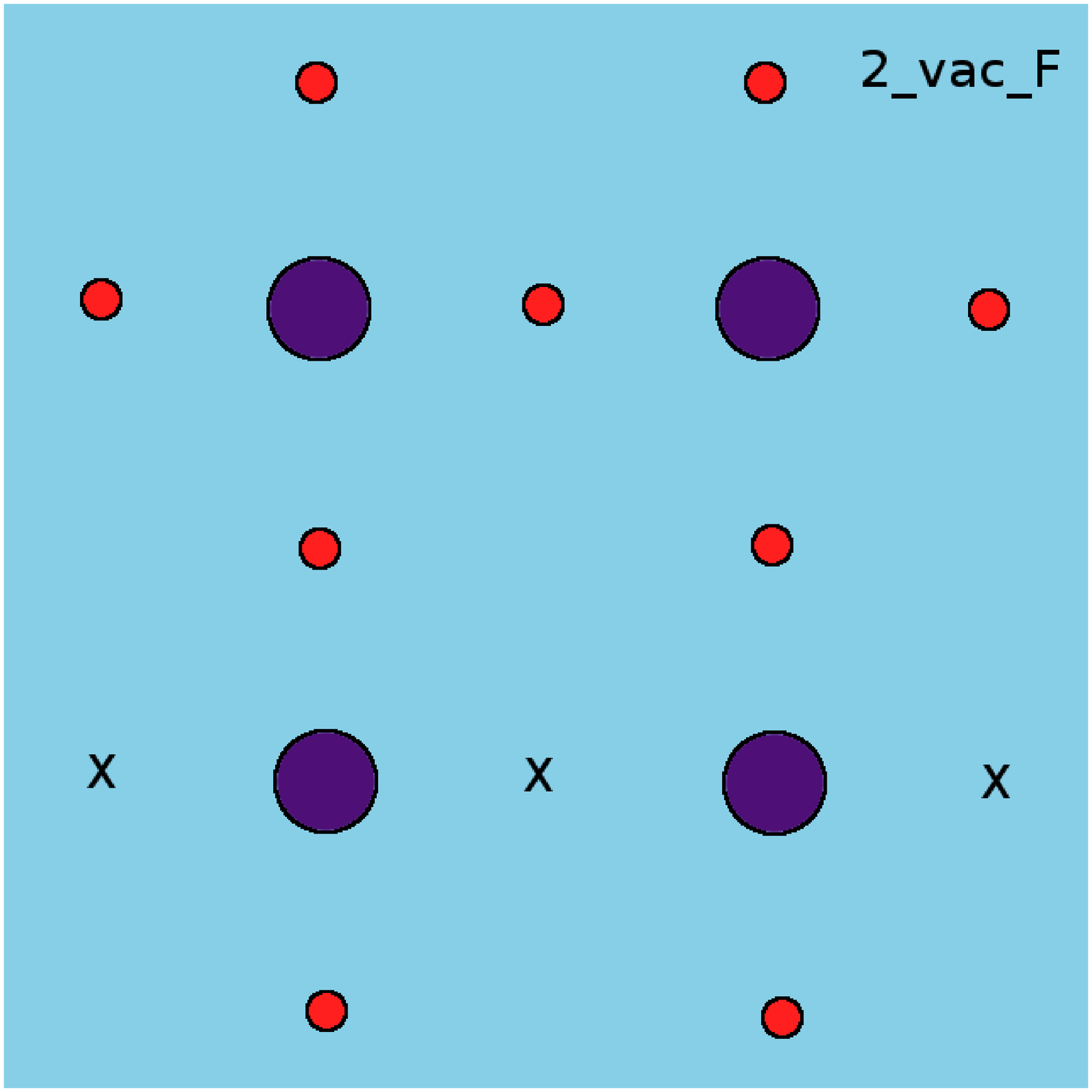}
\includegraphics[width=0.42\columnwidth]{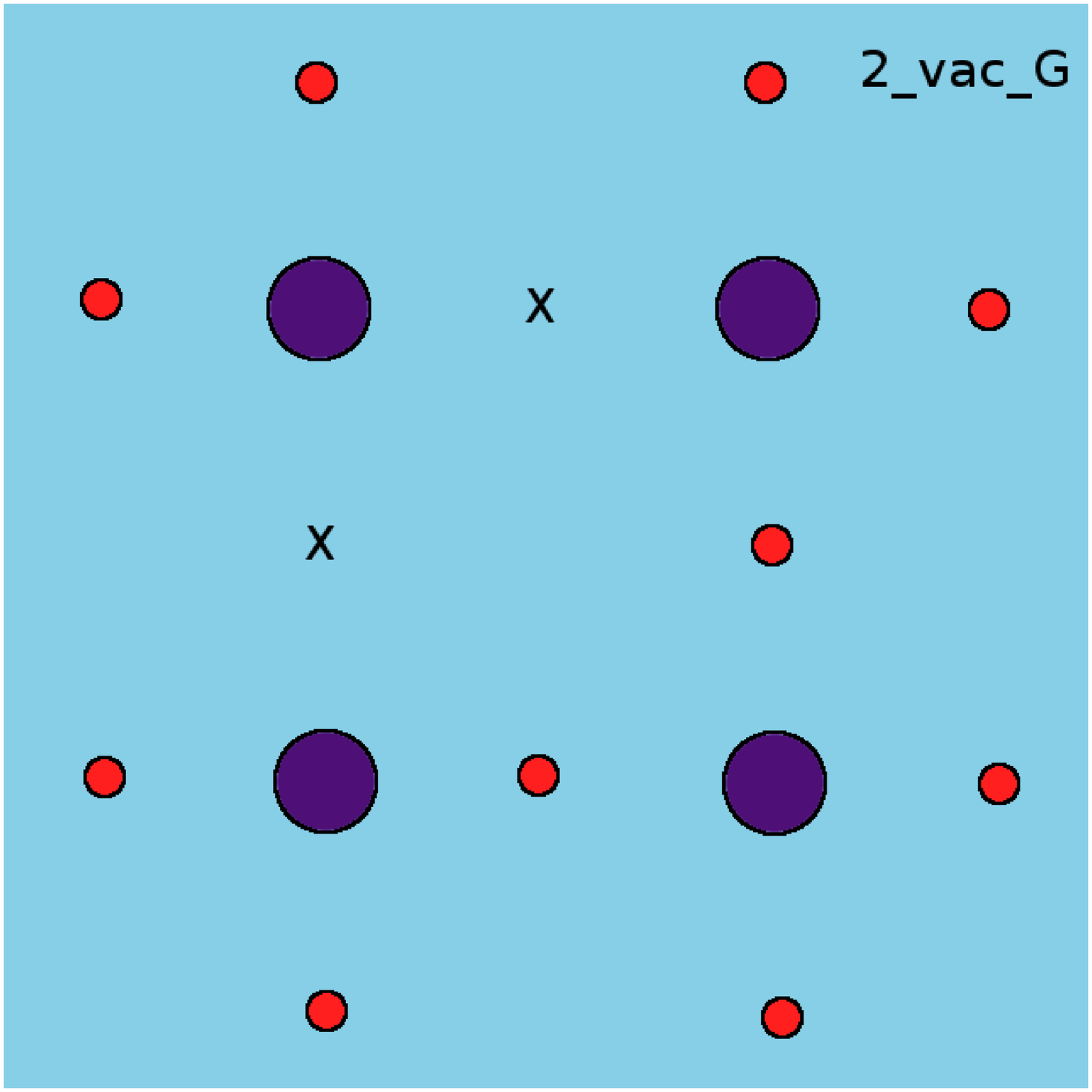}
\end{center}
\caption{(Color online.) Ten different oxygen-vacancy configurations, all of them projected on the plane described in Fig. \ref{unit_cell}. Co is in purple, O in red and the vacancies are represented by a cross. In each panel, the O of the left hand side are equivalent to those on the right and the bottom ones to the top ones.
0\_vac is the LCO without vacancies, 1\_vac\_A and 1\_vac\_B correspond to the two different possibilities when one O vacancy is introduced, 2\_vac\_A, 2\_vac\_B, 2\_vac\_C, 2\_vac\_D, 2\_vac\_E, 2\_vac\_F 
and 2\_vac\_G are the seven
two-O-vacancy configurations. The horizontal coordinate corresponds to the $b$ parameter and the vertical to the $c$.}
\label{O_vac_plane_pos}
\end{figure}

\section{Computational procedures}

Ab initio electronic structure calculations based on density functional theory (DFT)\cite{HK,KS} have been performed
using an all-electron full potential code ({\sc wien2k}\cite{WIEN2k}) on LCO.
The exchange-correlation term is parametrized depending
on the case:

We have used the generalized gradient
approximation (GGA) in the Perdew-Burke-Ernzerhof\cite{PBE} scheme for structural optimizations and to compute the energetics of the different configurations of LCO. 
These calculations were performed with a converged k-mesh and a value of  R$_{mt}$K$_{max}$= 5.0.
The $R_{mt}$ values used were in a.u.: 1.80 for La, 1.40 for Co and 1.11 for O when studying LCO energetics. The need to compare enegetics considering the O$_2$ molecule explains the choice of such small 
$R_{mt}$ values. In order to compute the energy of the O$_2$ molecule, an fcc lattice was used with the molecule at the center and a large enough unit cell so the bond-length is converged and it gives a value of 1.20 \AA, which compares reasonably well with the experimental O-O bond length of 1.21 \AA.\cite{web_NIST}

To include correlation effects in the Co d manifold, we have used the LDA+U method ($U=7$ eV, $J=0.7$ eV) in order to find a 
good description of the magnetic moments of the Co cations, their couplings and the overall electronic structure. A mapping of U between $4$ eV and $8$ eV was carried out to obtain the formation energy of the charged vacancies. 
A converged k-mesh and a value of  R$_{mt}$K$_{max}$= 6.0 were used.
The $R_{mt}$ values used were in a.u.: 2.33 for La, 1.83 for Co and 1.62 for O.

\section{Results and discussion}

We have divided the study in two main parts: structural characterization and electronic structure and magnetic properties of tetragonally strained LCO.
First, we need to understand how many oxygen vacancies will appear and how they are distributed. Once the structural ground state is found, we proceed to describe the electronic structure and magnetic properties that result from that ground state configuration.

\subsection{Structural characterization}

In this section of the analysis, we have characterized the structural properties of LCO when it has tetragonal symmetry and the in-plane lattice parameters 
are those of STO, mimicking the epitaxial strain effect introduced in an LCO film by the growth on an STO substrate. 
We have performed a systematic study that allows us to determine the optimal oxygen vacancy concentration as well as their distribution in the plane of vacancies. 
The unit cell that we have considered is shown in Fig. \ref{unit_cell}. Its stoichiometry is La$_{12}$Co$_{12}$O$_{36-n}$ where $n$ is the number of oxygen vancancies. 
We have fixed the lattice parameters $a$ and $b$ to reproduce the in-plane strain due to STO (3.905 \AA)\cite{prl_alex}. 
Parameter $c$ was left free for optimization, and so were optimized all the internal atomic coordinates. This $c$ direction is the (001) growth direction in the standard experiments with thin LCO films on a cubic substrate.

We have introduced one plane including O vacancies out of every 3 La-rich planes, as it is observed in experiments, such as Ref. \onlinecite{PRB_MVarela} (see Fig. \ref{unit_cell}, light blue plane). In this study, we have performed calculations for 10 different oxygen-vacancy 
configurations, as described in Fig. \ref{O_vac_plane_pos}. The number of possible configurations increases with the number of oxygen vacancies, that is one of the reasons why we stop the analysis in 2 vacancies. 
Also, with 3 vacancies or more we could not figure out a simple ionice model that would yield an insulating state at the same time as providing a magnetic moment close to the experimental value. However, we foresee that we have found a 2-oxygen-vacancy configuration 
that is in agreement with the experimental data. 
For choosing the different configurations we have taken into account the tetragonal symmetry of the system, considering that a basal O is different to an apical one, but retaining xy-symmetry for the analysis. 

In order to compute the energy $E$ of each configuration we have used the following equation: \cite{RevModPhys.86.253,PhysRevB.81.085212,Zhang_eq_charged_vac}

\begin{equation}\label{energy_conf}
 E=E_{conf}+\frac{n}{2}E_{O_2}-2nE_{e}-E_{0\_vac}
\end{equation}

where $E_{conf}$ is the energy of one of the configurations shown in Fig. \ref{O_vac_plane_pos} (the effect of charged vacancies is not included in this term). 
$E_{O_2}$ is the energy of the O$_2$ molecule, calculated as described in Section II, and $n$ the number of vacancies. $E_e$ is the term that provides the energy due to the charged vacancies. It is computed as the difference in energy 
between the top of the valence band of the configuration that we are calculating and the top of the valence band of the configuration without vacancies (0\_ vac). In order to set a common zero of energies for each 
case, we chose the energy of an O $1s$ orbital of an oxygen that lies away from the vacancy plane. The factor 2 is the charge, supposing they become fully ionized. 
Finally, $E_{0\_vac}$ is the energy of the configuration without vacancies 0\_vac, this term represents the zero of energies 
when obtaining $E$ from eq. (\ref{energy_conf}). We have assumed that all the vacancies form O$_2$ and we have also neglected the electrostatic interaction between the homogeneous background charge and the charged vacancies.

All terms in eq. (\ref{energy_conf}) were calculated using the GGA scheme, but $E_e$ was computed using the LDA+U scheme for various U values, since GGA leads to metallic solutions and the LDA+U method allows to compare the band alignment between two realistic insulating solutions, both for the stoichiometric and non-stoichiometric systems.

\begin{figure}[!ht]
\begin{center}
\includegraphics[width=\columnwidth]{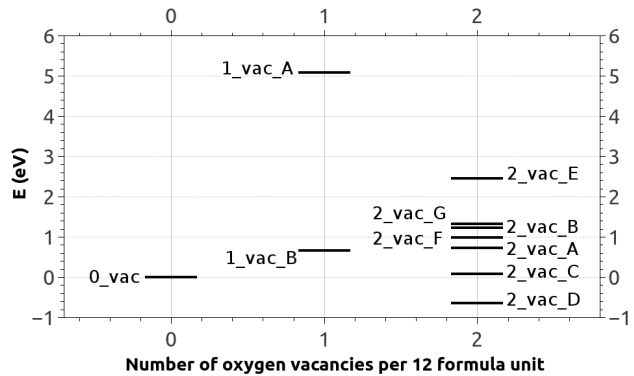}
\end{center}
\caption{Energetics for the different oxygen-vacancy configurations computed using eq. (\ref{energy_conf}). We can see that the most stable configuration is the 2\_vac\_D. 
This configuration corresponds to vacancy chains perpendicular to the (001) and contained in the vacancy plane.}
\label{energetics}
\end{figure}

The results for the energy of each oxygen-vacancy configuration using eq. (\ref{energy_conf}) are shown in Fig. \ref{energetics}. We observe that only configuration 2\_vac\_D is more stable than the configuration without vacancies. 
This corresponds to chains of oxygen vacancies perpendicular to the (001) and contained in the plane of vacancies (see Fig. \ref{O_vac_plane_pos}). 

Before analysing in detail 2\_vac\_D configuration, which is the ground state, we can explore the other configurations. For the case without vacancies, the system is a non-magnetic insulator. We also see that this configuration is 
quite stable compared to the others that present oxygen vacancies (2\_vac\_D is the only configuration with vacancies that is more stable than the stoichiometric solution). If we put an eye now on the two configurations with 
one vacancy, we will realize that one of them is much more stable than the other. The difference in energy between them is a sizable $0.37$ $eV/Co$. This result suggests us that in the case of adding a second vacancy, 
it could be energetically favorable to include it in an equivalent position to the one of 1\_vac\_B configuration, i.e, forming chains perpendicular to the (001) direction. We have demonstrated this statement calculating 
the energy for the 7 possible 2-vacancy configurations. As we have said 2\_vac\_D gives a stable solution. All the other $n=2$ cases are also FMI solutions, but higher in energy, except 2\_vac\_C, which is 
a ferromagnetic metal. We have tried several values of $U$ in a wide range, but an energy gap was not opened for that particular configuration.

\begin{figure*}[!ht]
\begin{center}
\includegraphics[width=16cm]{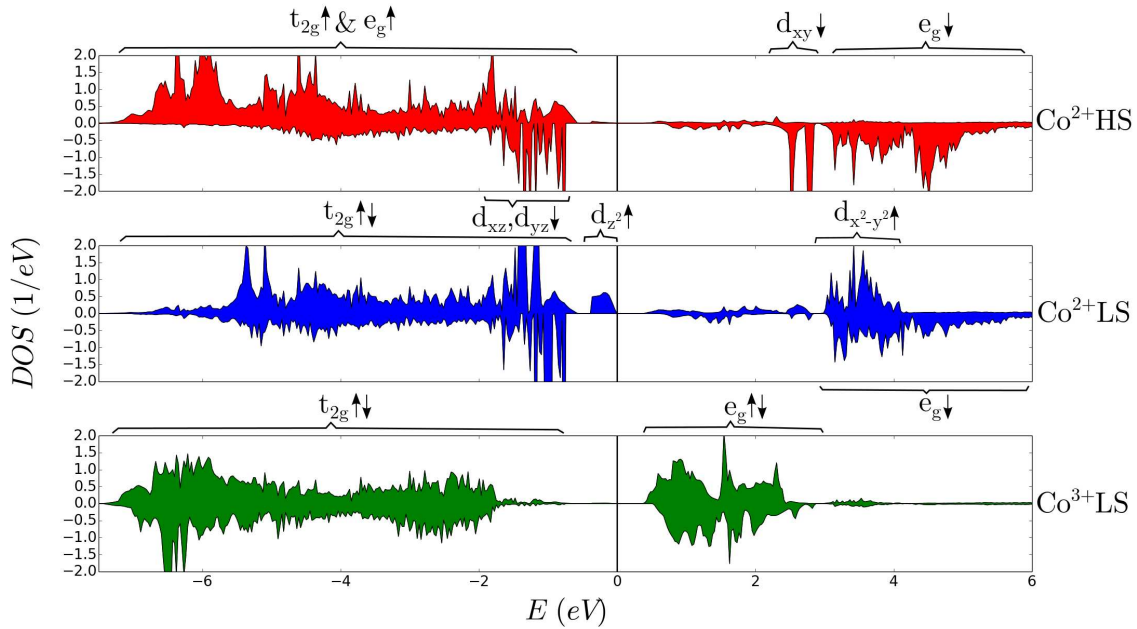}
\end{center}
\caption{(Color online.) Partial DOS for Co$^{2+}$ HS, Co$^{2+}$ LS and Co$^{3+}$ LS computed in the LDA+U scheme. The Fermi level is set to the zero of energies. Upper panel, Co$^{2+}$ HS: the $t_{2g}$ and the $e_{g}$ majority-spin states are occupied, and also the $d_{xz},d_{yz}$ minority orbitals. 
Middle panel, Co$^{2+}$ LS: the $t_{2g}$ majority and minority orbitals are occupied and so is the $d_{z^2}$ majority. Lower panel, Co$^{3+}$ LS: the $t_{2g}$ majority and minotiry orbitals are occupied.}
\label{DOS}
\end{figure*}

If we analyse in more detail the structure of the 2\_vac\_D configuration that we have obtained from our calculations, we find the following characteristics that can be compared with the experimental results 
reported in Ref. \onlinecite{PRB_MVarela}.
The lattice parameter along the (001) direction undergoes a contraction compared to the bulk value of LCO. We have obtained $c=3.810$ \AA, which is in good agreement with the experimental value $c=3.767$ \AA. \cite{PRB_MVarela}
The distance between La planes parallel to the plane of oxygen vacancies depends on whether one considers the distance between two La planes that contain the plane of vacancies or two stoichiometric planes. 
In the former case, with the plane of oxygen vacancies inside, we found that the distance is $4.02$ \AA. In the other case we found it is $3.85$ \AA. Experimentally, the same trend is observed but with a larger ratio. \cite{PRL_MVarela}

We have found in this subsection a configuration of oxygen vacancies that agrees with the experimental microscopy data obtained for LCO films grown on STO.

\subsection{Electronic structure and magnetic behavior}

In this section we analyse the electronic structure of the 2\_vac\_D configuration. As we have already said, this configuration presents a ferromagnetic-insulator behavior and is found to be a ground state. 
We will try first to explain the simple ionic model that accounts for the magnetic moments observed and its consistency with the ab initio calculations of the partial density of states (DOS) of the Co atoms shown in Fig. \ref{DOS}. 
In it, we can see the 3 types of Co atoms that appear in the converged solution: a non-magnetic Co atom six-fold coordinated and two magnetic Co atoms with 5-fold oxygen coordination, one with a higher value of the magnetic moment than the other.

\begin{figure}[!ht]
\begin{center}
\includegraphics[width=0.55\columnwidth]{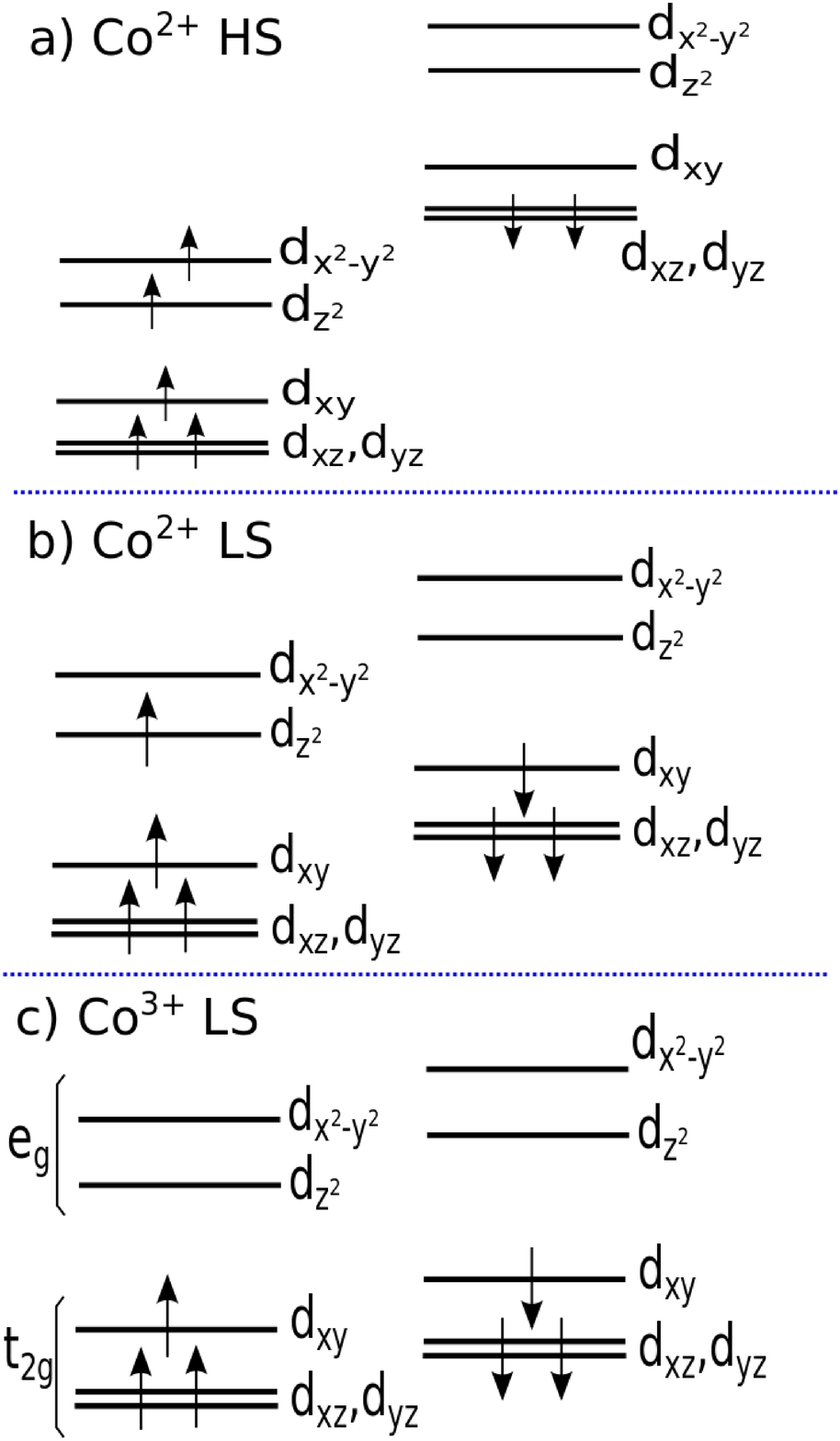}
\end{center}
\caption{Ionic model for Co$^{2+}$ HS, Co$^{2+}$ LS and Co$^{3+}$ LS. a) Co$^{2+}$ HS: the $t_{2g}$ and the $e_{g}$ majority states are occupied, and also the $d_{xz},d_{yz}$ minority orbitals. b) 
Co$^{2+}$ LS: the $t_{2g}$ majority and minority orbitals are occupied and moreover the $d_{z^2}$ majority. c) Co$^{3+}$ LS: the $t_{2g}$ majority and minority orbitals are occupied.}
\label{ionic_mod}
\end{figure}

From an ionic picture, if we consider the unit cell shown in Fig. \ref{unit_cell}, we have La$^{3+}$, O$^{2-}$ and Co$^{3+}$ for the stoichiometric compound without vacancies. The $d$ orbitals of Co$^{3+}$, which 
are in an octahedral environment, are 
occupied in the low-spin (LS) state, see Fig. \ref{ionic_mod} panel c), which gives rise to no-magnetic moment and an insulating state caused by crystal-field splitting between the Co t$_{2g}$ and $e_g$ levels that leads to a gap opening around the Fermi level in the simple diamagnetic configuration.

If we remove now two oxygens from the unit cell (Fig. \ref{unit_cell}), we have stoichiometrically La$_{12}$Co$_{12}$O$_{34}$. Structurally, there are four Co atoms that are five-fold coordinated and the remaining eight Co cations are six-fold octahedrally coordinated and remain non-magnetic, suggesting they retain the original 3+ valence. 
Our results disagree with previous spectroscopic evidences claiming Co$^{3+}$ high-spin (HS) atoms occuring in LCO strained films.\cite{PRB_MVarela} A simple electron count would naively imply that the other 4 Co atoms are Co$^{2+}$ cations. 
These are in a five-oxygen environment, that could be thought 
as an octahedral environment distorted in the z-axis. Such distortion breaks the degeneracy of the $t_{2g}$ and the $e_g$ levels as sketched in Fig. \ref{ionic_mod}. 
Our calculations show that the magnetic moment values are consistent with an ionic picture where half of these Co$^{2+}$ cations are in the LS state (in the 2\_vac\_D panel 
of Fig. \ref{O_vac_plane_pos} the cobalts on the left side), and the other half in the HS state (the ones on the right side). 
The actual values for the magnetic moments inside the muffin-tin spheres obtained from our calculations are of approximately 0.8 and 2.4 $\mu_B$, which account for the ionic values plus substantial hybridizations.
Taking into account Fig. \ref{ionic_mod}: for the case of Co$^{2+}$ LS (Fig. \ref{ionic_mod} b)), we observe that the $t_{2g}$ bands are fully occupied and there is one extra electron occupying the lower-lying $e_g$ state, which is of $d_{z^2}$ parentage due to the missing apical oxygen. 
Thus, the LS Co$^{2+}$ cations are in an $S=1/2$ state. 
In the case of Co$^{2+}$ HS (Fig. \ref{ionic_mod} a)), the majority channel is fully occupied, i.e., the $t_{2g}$ and the $e_{g}$ majority states are occupied, there are two minority $t_{2g}$ electrons occupying the lower-lying doublet $xz/yz$, which is split from the higher-lying $xy$ orbital due to the missing apical oxygen in the five-fold-coordinated environment. This leads to $S=3/2$ HS state. 
This simple ionic picture can be observed to be approximately reproduced (with substantial hybridizations, as is common in transition metal oxides) in the partial DOS for each inequivalent Co cation shown in Fig. \ref{DOS}. In this figure we show that the DFT calculations can be understood with the simplistic ionic model  
shown in Fig. \ref{ionic_mod}. The labels in Fig. \ref{DOS} help to see the correspondence with the ionic model.

The total spin magnetic moment in this solution is $0.67$ $\mu_B/Co$. If we now include spin-orbit coupling in the calculations, the total ordered moment obtained as the sum of $l_z+2s_z$ is about $0.8$ $\mu_B/Co$, since the magnetic Co cations present a non-negligible orbital angular momentum that becomes partially unquenched. This is in reasonable agreement with the saturation magnetization of the LCO thin films obtained experimentally, yielding a value of about $0.85$ $\mu_B/Co$. \cite{PRB_MVarela} 
This value approaches to the measured saturation magnetic moment, since the orbital momentum is parallel to the spin moment.

We also observe in Fig. \ref{DOS} that the energy gap that makes the system insulating is between the $d_{z^2}$ orbital of the Co$^{2+}$ LS (top of the valence band) and and the $e_g$ of the Co$^{3+}$ LS (bottom of the conduction band).
Of course the particular value of the gap will depend on the value of U chosen for the LDA+U calculations. In particular, for the DOS presented in Fig. \ref{DOS}, the value used was $7$ eV. 
According to our results, the insulating properties of these films occur naturally due to the additional band splittings introduced by non-octahedral environments. 

We have characterized in this section the electronic structure of the 2\_vac\_D configuration. We have related the calculated DOS to a simple ionic model that explains both the ferromagnetic ordering with a consistent value of the total magnetic moment and the insulating behavior.

\subsection{Stabilization of the ground state configuration with U}
\begin{figure}[!ht]
\begin{center}
\includegraphics[width=\columnwidth]{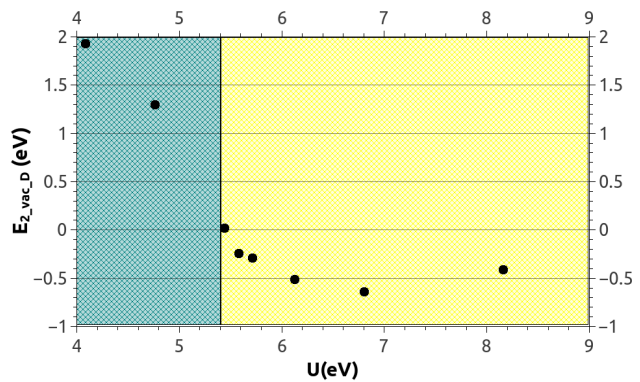}
\end{center}
\caption{(Color online.) Energy of the 2\_vac\_D configuration as a function of $U$ computed using eq. (\ref{energy_conf}). 
We observe that for values of $U>5.5$ $eV$ the configuration labeled 2\_vac\_D becomes stable.}
\label{2D_U}
\end{figure}

\begin{figure}[!h]
\begin{center}
\includegraphics[width=\columnwidth]{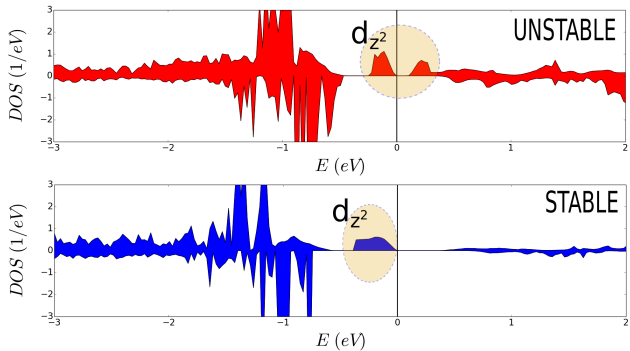}
\end{center}
\caption{(Color online.) Partial DOS Co$^{2+}$ LS for two different values of $U$. Upper panel: we observe that the $d_{z^2}$ orbital (highlighted in yellow) is not fully occupied which results in a solution that is not the ground state. 
Lower panel: we observe in this case that the $d_{z^2}$ orbital is fully occupied, which stabilizes the 2\_vac\_D configuration to become the ground state.}
\label{Co_SU}
\end{figure}

Regarding the value of $U$ used to compute the energies shown in Fig. \ref{energetics}, we have first performed calculations for different U values, and in particular Fig. \ref{energetics} was presented for $U=7$ eV. Figure \ref{2D_U} shows that there is a range of $U$ values where 2\_vac\_D becomes stable, and such U is within that range. This U value is similar to other values used in the literature for this kind of systems.\cite{LCO_spinstate_Blaha_Wentzcovitch} 
We observe that for values of $U$ less than $5.5$ $eV$ the 2\_vac\_D configuration is not stable. 
This can be analyzed through the change in the electronic structure that occurs when comparing the low-U and high-U solutions. Plots of the partial DOS of Co$^{2+}$ LS for two different values of $U$, one for an unstable case and another one for the stable case are shown in Fig. \ref{Co_SU}. 
We can see that the $d_{z^2}$ orbital of the Co$^{2+}$ LS is not fully occupied for the unstable case, while it becomes totally occupied for the stable case at larger U values. 
This change in the electronic structure is correlated with the stabilization of the 2\_vac\_D configuration.

\section{Concluding remarks}

In this study we have analysed the ferromagnetic-insulating behavior of LCO when it is grown on (001) STO. We have used DFT to analyze the electronic structure properties of LCO under strain and for various stoichiometries including different oxygen vacancy configurations and concentration.

We have found that the ground state of LCO when it is grown on STO is given by an off-stoichiometry of the form LaCoO$_{2.83}$ produced by about 6\% oxygen vacancies.  
We have shown that the vacancies form chains perpendicular to the (001) direction and are contained in the plane perpendicular to the film/substrate interface (the $ab$ plane in our calculations), consistent with experimental findings. Other structural features like the distance between La layers or the lattice 
parameter in the (001) direction agree with the experimental measurements.

We found that the Co atoms that lie in the plane of vacancies are a mixture between  
Co$^{2+}$ LS and Co$^{2+}$ HS. The total magnetic moment is in close agreement with experimental measurements. 
The ferromagnetic insulating behavior of the film is readily obtained through a simple ionic model, 
the gap opening occuring naturally in that electron count due to the appearing crystal field splittings together with the addition of a reasonable U value. 

In conclusion, we can say that the strain introduced by STO on LCO favors the presence of chains of oxygen vacancies perpendicular to the (001) direction. Ferromagnetism arises from the inclusion of those vacancy chains and can be explained with an ionic model.
The results presented in this work could lead to a better understanding and desingning of other ferromagnetic-insulator oxides in which oxygen vacancies could play an important role.

\acknowledgments

This work has been supported by the MINECO of Spain through the project MAT2016-80762-R.

\end{document}